\documentclass{iopjournal}

\usepackage{multirow}
\usepackage{cite}

\begin{document}

\articletype{Article type} 

\title{Hybrid films of Co – C${_{60}}$ preparation and changes induced by external stimuli.}

\author{Giovanni Ceccio$^{1,*}$\orcid{0000-0001-7959-8815}, 
Jiri Vacik$^1$\orcid{0000-0002-1875-4914},
Yuto Kondo$^2$,
Kazumasa Takahashi$^2$\orcid{0000-0001-7872-0225},
Romana Miksova$^1$\orcid{0000-0002-8887-601X},
Eva Stepanovska$^{1,3}$\orcid{0000-0003-1973-6539},
Josef Novak$^{1,3}$\orcid{0000-0002-1231-4619},
Petr Malinsky$^{1,3}$\orcid{0000-0002-4236-1192},
Barbara Fazio$^4$\orcid{0000-0002-1947-1123},
Catia Cannilla$^5$\orcid{0000-0001-6773-3784},
Alena Michalcova$^6$\orcid{0000-0002-1225-5380},
and Sebastiano Vasi$^{7,8,*}$\orcid{0000-0003-0480-3321}}

\affil{$^1$Department of Neutron and Ion Methods, Nuclear Physics Institute of Czech Academy of Science, Hlavní čp. 130 Husinec - Řež, 25068, Czech Republic}

\affil{$^2$ Department of Electrical Engineering, Nagaoka University of Technology, 1603-1 Kamitomioka, Nagaoka, Niigata, 940-2188, Japan}

\affil{$^3$  Department of Physics, Faculty of Science, J. E. Purkyne University in Ústí nad Labem, Pasteurova 3632/15, 400 96 Ústí nad Labem, Czech Republic}

\affil{$^4$ IMM-CNR sede di Messina, Viale Ferdinando Stagno d’Alcontres, 31, 98166, Messina, Italy}

\affil{$^5$ CNR-ITAE, Istituto di Tecnologie Avanzate per l’Energia, Via S. Lucia Sopra Contesse 5, 98126, Messina, Italy}

\affil{$^6$ Department of Metals and Corrosion Engineering, University of Chemistry and Technology in Prague, Technicka 5, Prague 6, 166 28, Czech Republic}

\affil{$^7$  Department of Mathematics and Computer Sciences, University of Messina, Physical Sciences and Earth Sciences, Viale F. Stagno d’Alcontres 31, 98166, Messina, Italy}

\affil{$^8$  OPENFIS S.R.L. - SPIN OFF ACCADEMICO, Viale F. Stagno D’Alcontres 31, 98166, University of Messina, LABORATORIO A2AT3, Messina, Italy}

\affil{$^*$Author to whom any correspondence should be addressed.}

\email{ceccio@ujf.cas.cz}


\begin{abstract}
In this work, we report on the study on organic-metal hybrid systems, in particular Co-C$_{60}$ fullerene thin films. This study mainly focused on the investigation of the morphological and structural evolution of the film surface after various external stimuli designed to provide energy to the system. For film growth, we adopted an innovative approach, combining ion‑beam sputtering of a pure metal target with thermal evaporation of C$_{60}$ in a co‑deposition setup. The films underwent a series of treatments to induce modifications. Laser and ion irradiations were performed using a pulsed laser, a continuous Ar beam, and a pulsed C beam. In addition, thermal annealing in vacuum was performed to examine the long-term effects of temperature.
The composition of deposited film was investigated using Ion Beam Analysis, the morphology and the structure, and the effects of treatments on the films were studied using SEM and TEM microscopies and Raman spectroscopy. Changes in electrical resistance were also measured to explore potential applications of these films after treatment.  
\end{abstract}

\section{Introduction}
Recent advances in nanoscience, particularly in nanoarchitectures, highlight metal–organic hybrid nanostructures as an increasingly relevant research topic, yielding intriguing results in electronics and proposing advanced applications, such as catalysis and electrochemistry, energy storage and conversion, gas sensing and spintronics \cite{Scheele_2015, Goiri_2015, Cui_2016, Devkota_2016, Otero_2017, Sosa_2018, Song_2019, Socol_2020, Shao_2023, langer2022graphene,lin2021doping, wang2018situ,deng2016catalysis, zhang2020single, cortes2018fe,zhou2011adsorption}. 
The study of C$_{60}$ fullerene in such hybrid systems is often taken into consideration due to the unique properties of C$_{60}$ molecules, such as the exciting atomic structure, the intriguing coupling with metals, the tiny hyperfine interactions, the greater structural stability and the relative low cost. These features collectively increase the interest in such materials \cite{Tanaka_2002, Konarev_2003, Yoshikawa_2009,vishnoi2020optical, karczmarska2022carbon, joshi2021fabrication}.
Among various C$_{60}$-based hybrid systems, self-assembled nanostructured materials are particularly notable for their tunability, offering precise control over the nanostructure during growth or post-production to tailor their electrical properties. The discovery of higher conductivity and superconductivity in alkali metal fullerides has also been a strong motivation for the implementation of transition metal-fullerene related research \cite{varshney1999superconductivity, takeya2013preparation,vishnoi2020optical,Vacik_2010}. Attempts to combine transition metals (TM) and C$_{60}$ via co-deposition to access TM-fullerides have usually revealed pronounced metal clustering (due to the high cohesive energy of metals) with separation from C$_{60}$ phase, which hinders homogeneous mixing but also enables additional functional behaviour of such nanostructures \cite{Vacik_2010, Bolokang_2012}. 
An example is given by structures based on the coexistence of Co and C$_{60}$, in both homogeneous compounds and heterogeneous films, which have a great impact in spintronics due to the catalytic and magnetic properties of Co \cite{Tanaka_2002, Kaushik_2022, Gupta_2023, Patel_2023, lee2002excellent, kuznetsov2012magnetic}. Furthermore, such hybrid systems have gained interest for potential applications in lithium-ion batteries \cite{ceccio2024study}, where structural stability, conductivity, and controlled phase evolution are critical for improving performance and longevity. Besides their interesting properties, however, integrated hybrid systems are considered thermodynamically and structurally unstable (mainly due to the high internal stress resulting from the mixing of mostly immiscible phases and the vulnerability of C$_{60}$ molecular cages due to their easy photo-oxidation, polymerization or fragmentation in an environment exhibiting strong catalytic properties) \cite{vacik2016laser, zalibera2021metallofullerene, vacik2009hybridization}. This makes their possible applications more challenging. The use of various disruptive agents (such as thermal annealing, ion irradiation, laser illumination, chemical reagents, etc.) can significantly influence thermodynamically unstable systems, inducing their structural changes at the nanoscopic and macroscopic scales as the system moves toward a state of minimum energy. This approach not only enables control over material properties for specific applications, but also serves as a model for studying the long-term evolution of these hybrid systems under extreme conditions, which is highly relevant for their potential integration into devices designed for harsh environments (e.g., space applications, high-radiation areas, and extreme temperature conditions).
While numerous studies have investigated the growth, structure, and intrinsic properties of metal–C$_{60}$ hybrid systems, most reported works focus on as-deposited films or on single modification routes, often produced under different deposition conditions or with varying compositions \cite{sakai2007comparative,Vacik_2010,Kaushik_2022}. As a result, a direct and systematic comparison of how different external stimuli influence the same Co-C$_{60}$ system remains limited \cite{vishnoi2020optical}.
The ability to intentionally modify the system provides the advantage of preventing unwanted changes due to system aging or adverse environmental conditions during the lifetime of a possible device based on such systems. Our previous systematic works \cite{lavrentiev2020structure, lavrentiev2021tuneable,lavrentiev2024room} have focused on study the physical properties of systems realized using different ratios of constituents within thin films. In contrast to these previous studies, the present work does not aim to explore new compositions or mixing ratios, but rather to isolate the effects of different post-deposition energy inputs applied to films fabricated under strictly identical growth conditions. This post-treatment–centered approach allows us to disentangle how distinct excitation pathways drive structural reorganization and functional property changes in the same hybrid system. The prepared samples have been exposed to vacuum annealing, pulsed laser irradiation, and ion irradiation (continuous beam and pulsed beam). The analyses performed show that the hybrid systems evolved in different ways, exhibiting diversification in their nanostructures and macrostructures depending on the external perturbation applied.

\section{Materials and Methods}

\begin{figure}[t]
\centering
\includegraphics[width=0.8\textwidth]{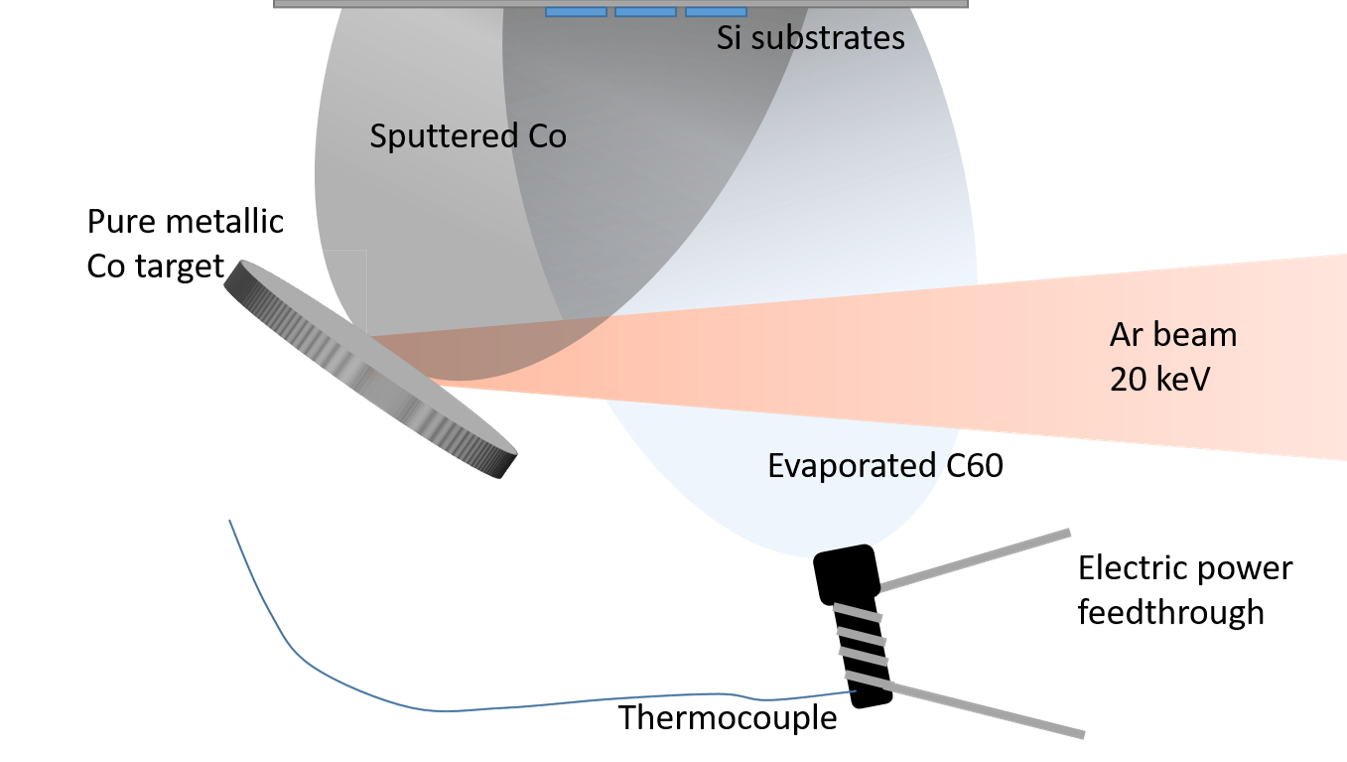}
\caption{Deposition scheme with IBS setup and thermal evaporator}
\label{fig1}

\end{figure}

We have produced a set of Co - C$_{60}$ mixture films by co-deposition of Co and C$_{60}$ on Si(110) substrates in high vacuum (base vacuum, 2E-6 mbar). The growth of the Co film was performed by the ion beam sputtering technique realized by using a Low Energy Ion Facility (LEIF) \cite{ceccio2024study, vacik2024study}, operating with a beam of $Ar^+$ ions accelerated at 20 keV and focused on a Co target (2 inch x 0.125 inch, 99.99\% purity, Kurt J. Lesker) mounted on a suitable target holder and inclined at 45 degrees with respect to the beam. The target faces the substrates mounted at a distance of 100 mm, under an angle of 30 degrees. Just below the substrate holder, at the distance of 100 mm, a lab made evaporator that acts as a source of C$_{60}$ (99.9\% purity, Nanografi) is present, in order to perform simultaneous deposition of Co and C$_{60}$ (see sketch if Fig.\ref{fig1}). The evaporator, equipped with digital temperature control and a thermocouple inserted directly into the crucible, was maintained at a temperature of 450 $^\circ$C for the entire duration of the deposition. During the beam calibration and heat up procedure of the crucible, a shutter is present to avoid unwanted depositions on the substrates. For TEM analytical purpose, samples were deposited on 300 mesh Cu TEM grids with holey carbon films (TedPella, Inc.). 

All films were deposited together in order to guarantee the same experimental parameters and the same film on all substrates.  Ion Beam Analysis experiments were performed at the Tandetron MC 4130 tandem accelerator located at NPI. RBS and ERDA were used for the analysis. To identify heavier elements, $He^+$ ions with an energy of 2 MeV were used, while for the carbon detection $H^+$ ions at 1.735 MeV were chosen, hitting the films with an incident angle of 7$^\circ$). The backscattered ions were detected using an ORTEC ULTRA-series detector, which features a 50 mm² active area and a 300 $\mu$m thick depletion layer. The detector was positioned at a backscattering angle of 170$^\circ$ out of the plane, in accordance with Cornell geometry. 
In the ERDA measurement, $He^+$ ions with an energy of 2.5 MeV were utilized to identify H, and the detector was placed at a backscattering angle of 30$^\circ$ in the plane, following the IBM geometry. The recoiled particles were detected by a Canberra PIPS detector preceded by a 12 $\mu m$ thick Mylar foil. 
Data obtained by ion beam analysis were processed using the SIMNRA code \cite{mayer1999simnra}.
The irradiation of the samples by means of a continuous beam of $Ar^+$ was carried out using the LEIF (NPI) system, positioning the samples orthogonal to the beam in place of the sputtering target. For irradiation, an energy of 20 keV and a current of 500 $\mu$A were used up to a total fluence of 1E15 Ar/cm$^2$. Another sample set was irradiated at the same fluence and energy using a pulsed $C^+$ beam. A Laser Ion Source (Nagaoka University) was used to produce a pulsed C beam. The used laser source was a Quantel Brillant operating at 10Hz repetition rate, 6ns pulse duration, 532 nm wavelength and laser energy of 130 mJ . The laser pulse was focused (in vacuum) on the surface of a C target held at a voltage of 20kV and facing the grounded samples. To reach the same fluence used for the Ar beam, a total of 12000 pulses were needed for each sample. An additional set of samples was exposed directly to the laser source, to induce changes on the surface. The laser irradiations occurred in air, without laser focusing and at an energy of 5 mJ per pulse. Each sample was irradiated with 1000 pulses. The annealing was performed using a special vacuum chamber with a heating system inside. The annealing was performed in a vacuum level of 1E-5 mbar and for a duration of 5h (without considering the cooling time) at temperature of 300$^\circ$C.
The samples, pristine and modified, were investigated for morphological evolution by scanning electron microscopy SEM Hitachi-SU8230. Additional morphological study with elemental mapping of the materials was carried out using the Thermo Scientific Helios™ 5 UC Dual Beam Scanning Electron Microscope (SEM-FEG-UHR) equipped with different detector ETD (Everhart-Thornley Detector), TLD (Through-the-Lens) and STEM (Scanning Transmission Electron Microscopy). The elemental analysis and mapping were carried out by using the energy dispersive analysis system (EDX). 
Measurements were carried out with an operating voltage between 10 and 30 kV, without pretreatment of the samples.	
The TEM analysis of microstructure was performed by use of Jeol 2200 FS (Jeol,Tokio, Japan) field emission gun (FEG) equipped with with an Oxford Instruments EDS analyzer, instrument operating at a 200 kV accelerating voltage with a point resolution of 2.4 Å. The micrographs have been carried out using a TVIPS camera and EM-Menu software. 

Raman spectroscopy measurements were also performed, using a LabRam HR-EVO Horiba operating at 532 nm, 50X LWD with power 0.8 mW per $\mu m^2$ and CCD Syncerity Horiba and NRS 7200 microRaman spectrometer with 532 nm wavelength and applied power 6.5 mW. In particular, the micro-Raman spectrometer was used for the evaluation of the as-deposited and modified films, on the other hand the Raman spectrometer was used for the fullerene characterization. 
Changes in electrical resistivity of the pristine and modified samples, were investigated using 2182A Nanovoltmeter and 6221 DC current source Keithley on the surface of sample, in a galvanostatic setup. The measurements were performed using the standard two-point method in air and room temperature. While the two-point method includes contact and lead resistances, all measurements were performed under identical conditions, allowing reliable comparison of relative resistance changes induced by different treatments.

\section{Results}
\begin{figure}[!ht]
\begin{minipage}[h]{0.5\linewidth}
\begin{center}
\includegraphics[width=1\linewidth]{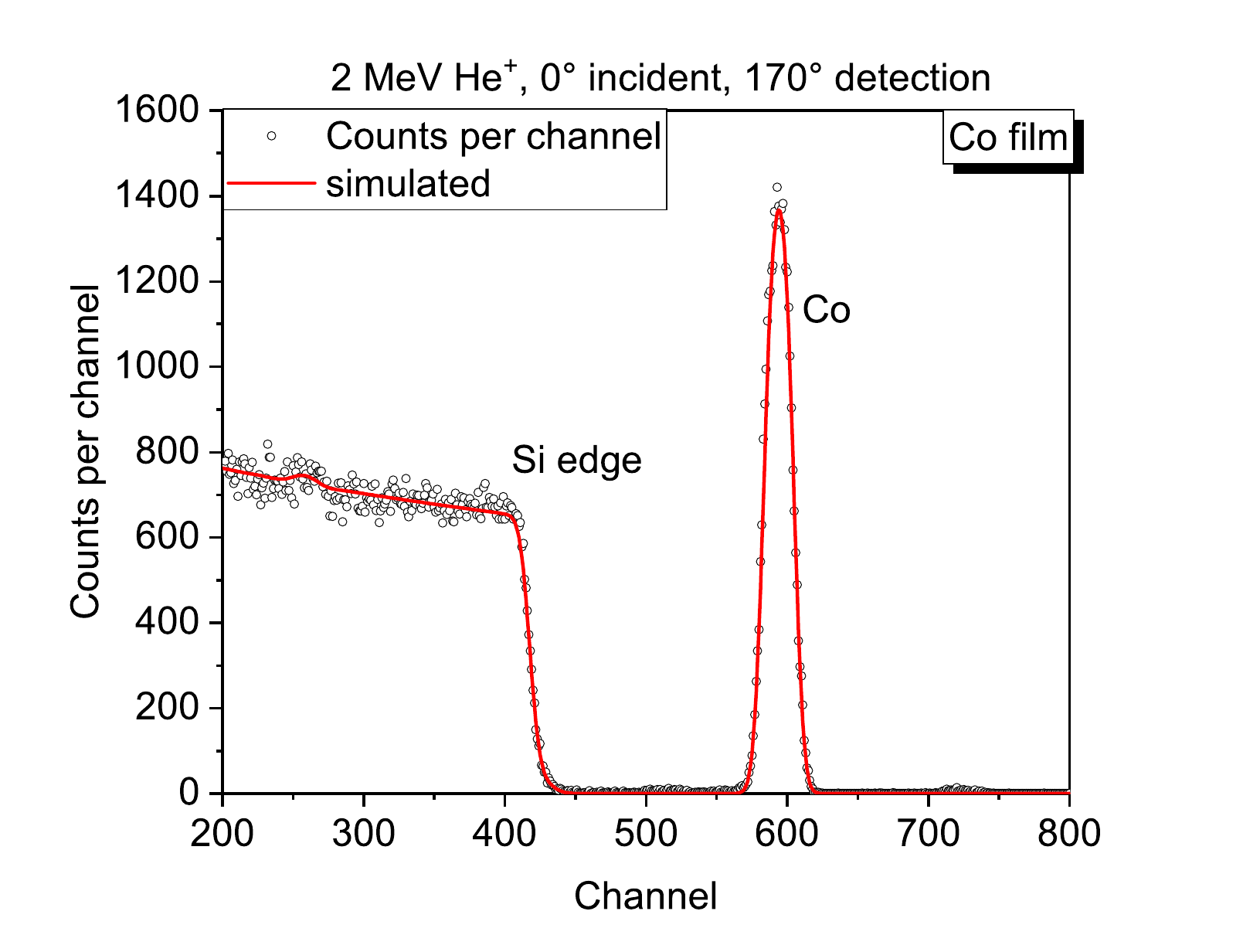} 
\label{fig2a}
\end{center} 
\end{minipage}
\begin{minipage}[h]{0.5\linewidth}
\begin{center}
\includegraphics[width=1\linewidth]{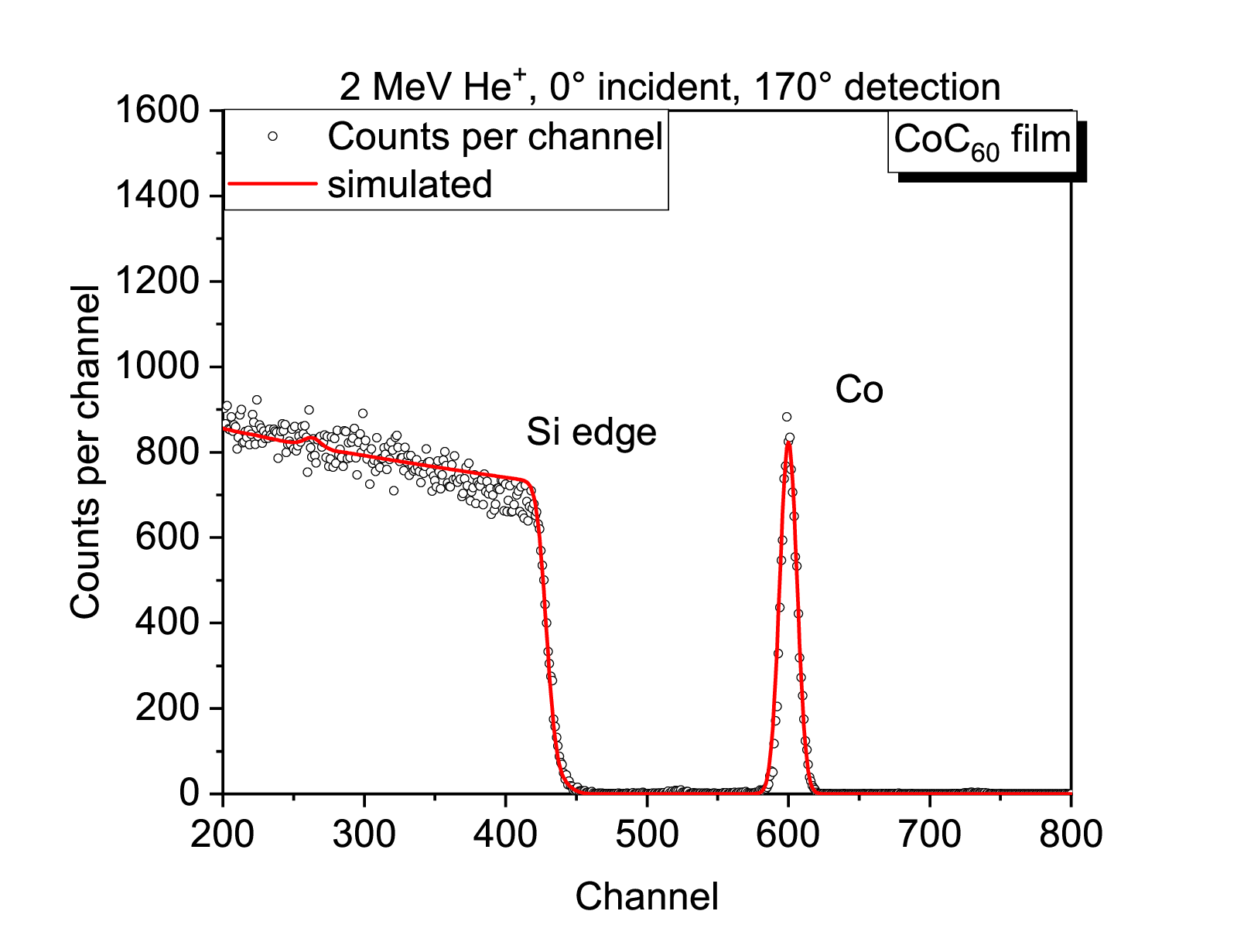} 
\label{fig2b}
\end{center}
\end{minipage}
\caption{RBS results on pure Co and Co - C$_{60}$ samples obtained with 2MeV $He^+$. Black dots are experimental data and the red line represents the fit obtained from SIMNRA simulation.}
\label{figure_rbs}
\end{figure}

The chemical composition of the produced films was investigated after deposition using Rutherford Backscattering Spectrometry (RBS) and Elastic Recoil Detection Analysis (ERDA) techniques. These scattering techniques are used for compositional thin film analysis, based on stopping power and inelastic energy loss during ion-solid interactions and the kinematically recoil. In our case, RBS is particularly suitable for the detection of heavy Co atoms (because light C atoms are hidden by Silicon signal), meanwhile the ERDA is used for H detection. Experimental results relative to 2 MeV $He^+$ are shown in Fig. \ref{figure_rbs} with simulated fit. Note that the spectra were normalized to the substrate. Carbon measurements were performed using 1.735 MeV $H^+$ ions and different incident ions angle (0 degrees were used for measurement with $He^+$ analysis and 7 degrees for the $H^+$). The relative spectra are illustrated in Fig. S1 of the Supplementary information (SI). Table \ref{tab1} includes all the elemental concentrations' values of the data acquired by performing RBS and ERDA analyses on Co and Co - C$_{60}$ films. RBS measurements confirm that CoC$_{60}$ hybrid films have been successfully fabricated.  
        \\
\begin{table}[h]
\caption{Elemental composition of samples calculated by RBS-ERDA analysis using both 2MeV $He^+$ beam and 1.735 MeV $H^+$.}
\centering
\resizebox{10cm}{!} {
\begin{tabular}{c c c c c c}
\hline
\multirow{2}{*}{Sample} & Co & C & O & H & Thickness \\
 & [at. \%] &	[at. \%] &	[at. \%] &	[at. \%] &	[10$^{15}$ at/cm$^2$] \\
 \hline
Co film &	70 &	14 &	5 &	11 &	480 \\
\hline
Co - C$_{60}$ film &	50	& 33	& 3 &	14	& 245 \\
\hline
\end{tabular}
}

\label{tab1}
\end{table}

Oxygen and carbon contaminations as well as hydrogen content are detected in the produced films. The sample with C$_{60}$ show almost a ratio 3:2 between Co and C atoms. It should be noted that the pure Co film was prepared and analyzed exclusively as a compositional and structural reference for ion beam analysis. The electrical, morphological, and post-treatment comparisons discussed in this work are performed only among Co–C$_{60}$ hybrid films, which share the same deposition conditions and comparable thickness. Therefore, the difference in areal density between pure Co and Co–C$_{60}$ films does not introduce bias in the comparative interpretation of the hybrid film properties.

Presence of oxygen in the deposited films can be explained as oxidation of the Co NPs (possibly formed in the film nanostructure) during the air exposure of the samples. \\      
The quantity of oxygen is independent of the presence of C$_{60}$ or the ratio between Co and C$_{60}$. This may suggest that the detected oxygen is bounded chemically to the cobalt, forming more realistic composition $Co_xO_yC_{60}$. By coming into contact with the Co clusters, the O$_2$ molecules react with the Co surface causing O$_2$ dissociation and formation of the $Co_xO_y$\cite{klingenberg1993oxygen}. By considering that the presence of C atoms in pure Co film (due to vacuum contamination by partial sputter of holder and shutter probably) and Co-C$_{60}$ film are 14\% and 33\%, respectively, we can confirm that the C$_{60}$ is almost the 20\% of the film, so the majority of the film consists of Co atoms. 
As illustrated in Fig. S2 of SI, the RBS analysis allows us to also establish that the distributions of Co and C, as well as the other elements, within the thickness of the samples are constant.  
Furthermore, thanks to the ERDA results, the presence of about 10\% of H atoms was measured and confirmed as shown later by Raman analysis. 
\begin{figure}[!ht]

\begin{center}
\includegraphics[width=0.4\textwidth]{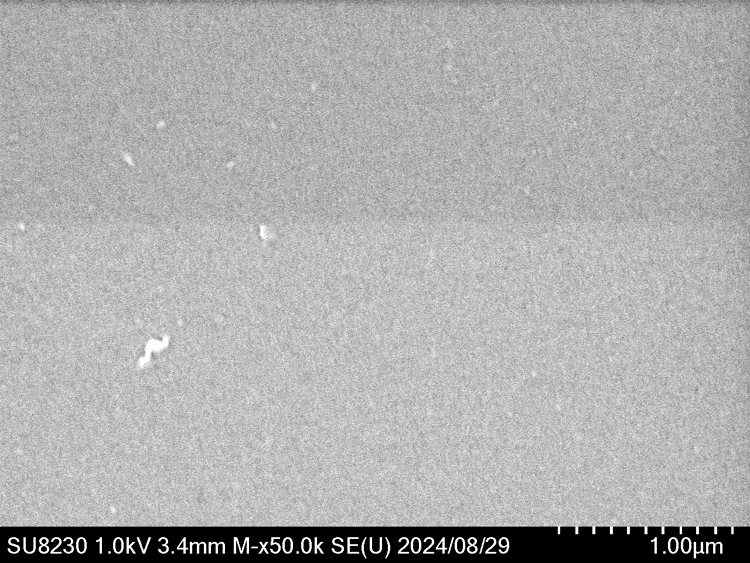}

\end{center}
\caption{SEM pictures taken at 50kX magnification of as deposited sample.}
\label{fig_SEM_as_prepared}
\end{figure}
\begin{figure}[!ht]

\begin{center}
\includegraphics[width=0.8\textwidth]{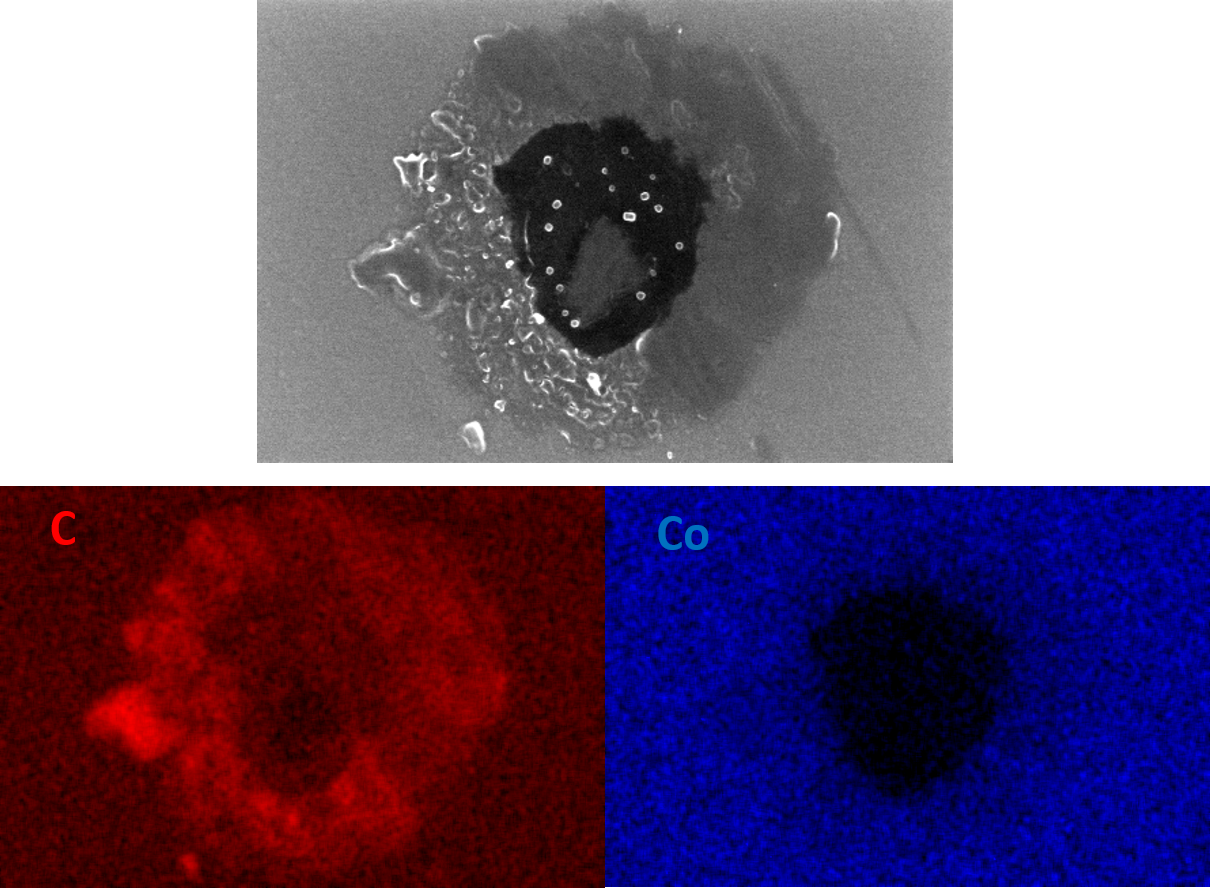}

\end{center}
\caption{Elemental mapping of particle on the as deposited sample.}
\label{pristine_mapping}
\end{figure}

\begin{figure}[!ht]
\begin{center}
\includegraphics[width=0.4\textwidth]{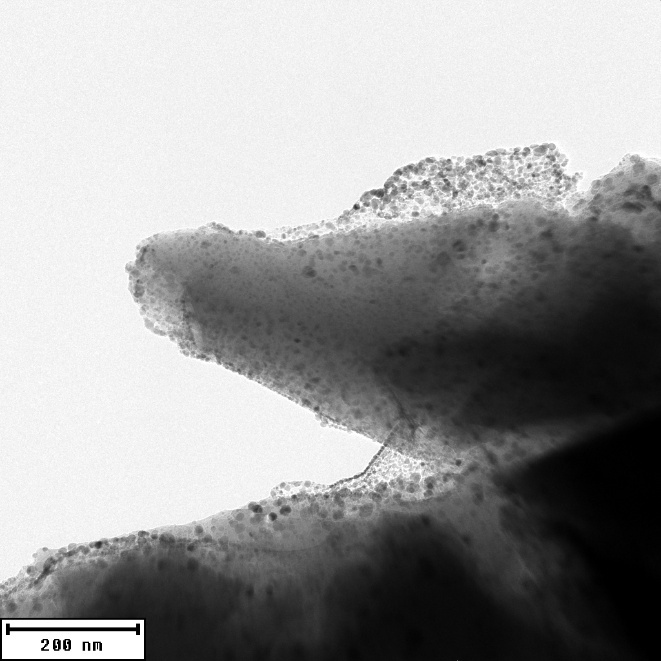}

\end{center}
\caption{TEM micrograph of as prepared sample.}
\label{TEM_pristine}
\end{figure}

\begin{figure}[!ht]

\begin{center}
\includegraphics[width=0.8\textwidth]{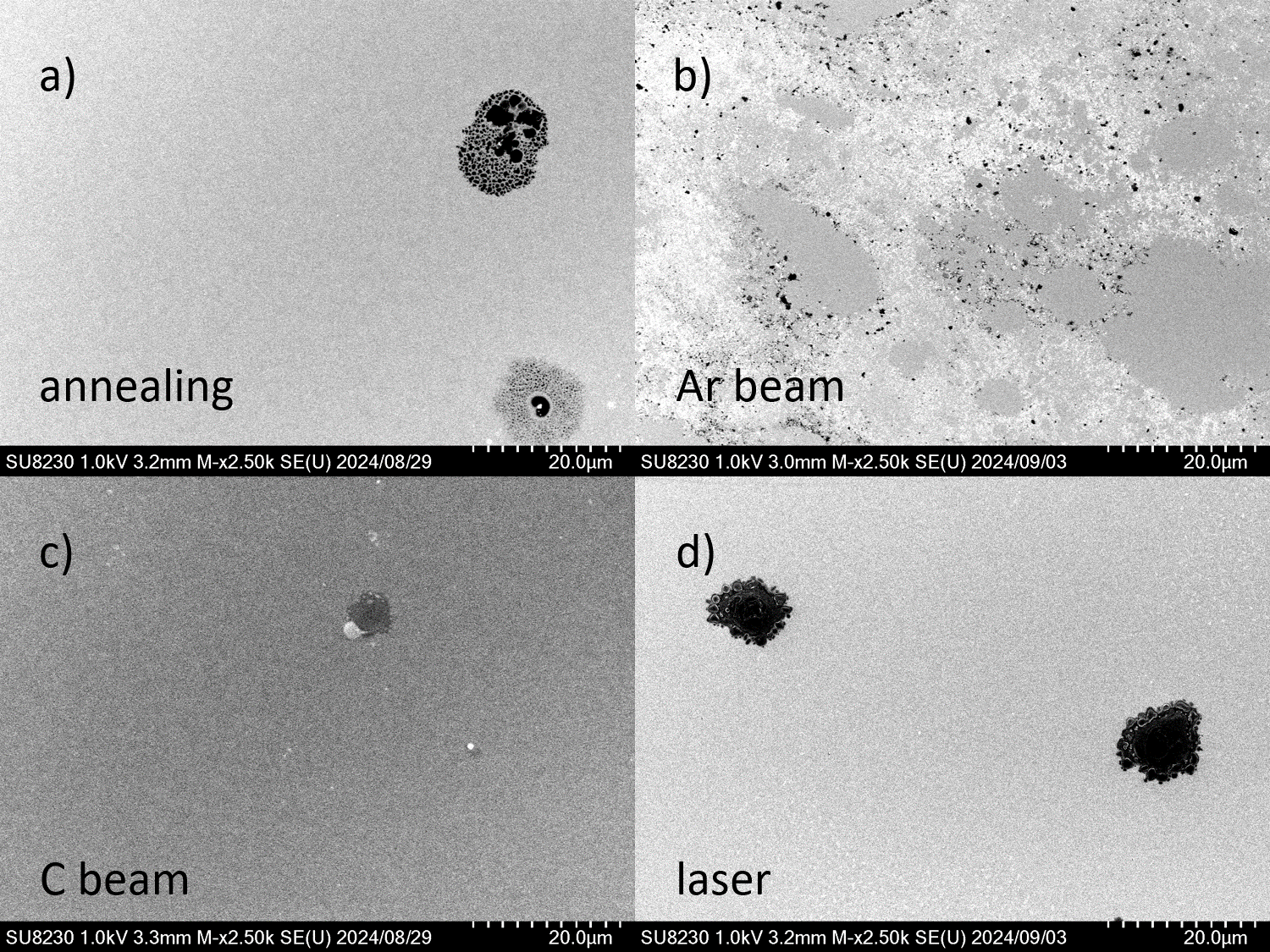}

\end{center}
\caption{SEM pictures taken at 2.5kX magnification modified samples.}
\label{fig_SEM}
\end{figure}

Fig.\ref{fig_SEM_as_prepared} reports the SEM images for the as prepared samples and Fig.\ref{pristine_mapping} shows elemental mapping of selected particles obtained using EDX analysis. In the SEM micrograph of pristine sample, smooth surface and uniform particles distribution are recognized, indicating homogeneous mixture with few visible micron sized particles. 
EDX analyses reported in Fig.\ref{pristine_mapping} were performed on a selected particle to understand its composition. It was found a deficiency of Co inside the particle with a Co-enriched ring around it, and uniform distribution anywhere else.
In order to observe the formed nanostructures in the hybrid CoC$_{60}$  films, we performed detailed TEM analyses on the as deposited film. So, following the procedure reported in the “Materials and Methods” section, we deposited the sample on Cu holey carbon grid to analyze it for structural information. Fig.\ref{TEM_pristine} shows the TEM image obtained from the CoOC$_{60}$ film. The sample has composite nanostructure consisting of small Co nanoparticles distributed uniformly.
Samples subject to modification were analyzed  from morphological point of view as the pristine one. The SEM images from the modified samples are reported in Fig.\ref{fig_SEM} taken at a magnification of 2.5kx. For each of these samples was performed as well elemental mapping with EDX analysis on selected particles.

\begin{figure}[!ht]

\centering
\includegraphics[width=.8\textwidth]{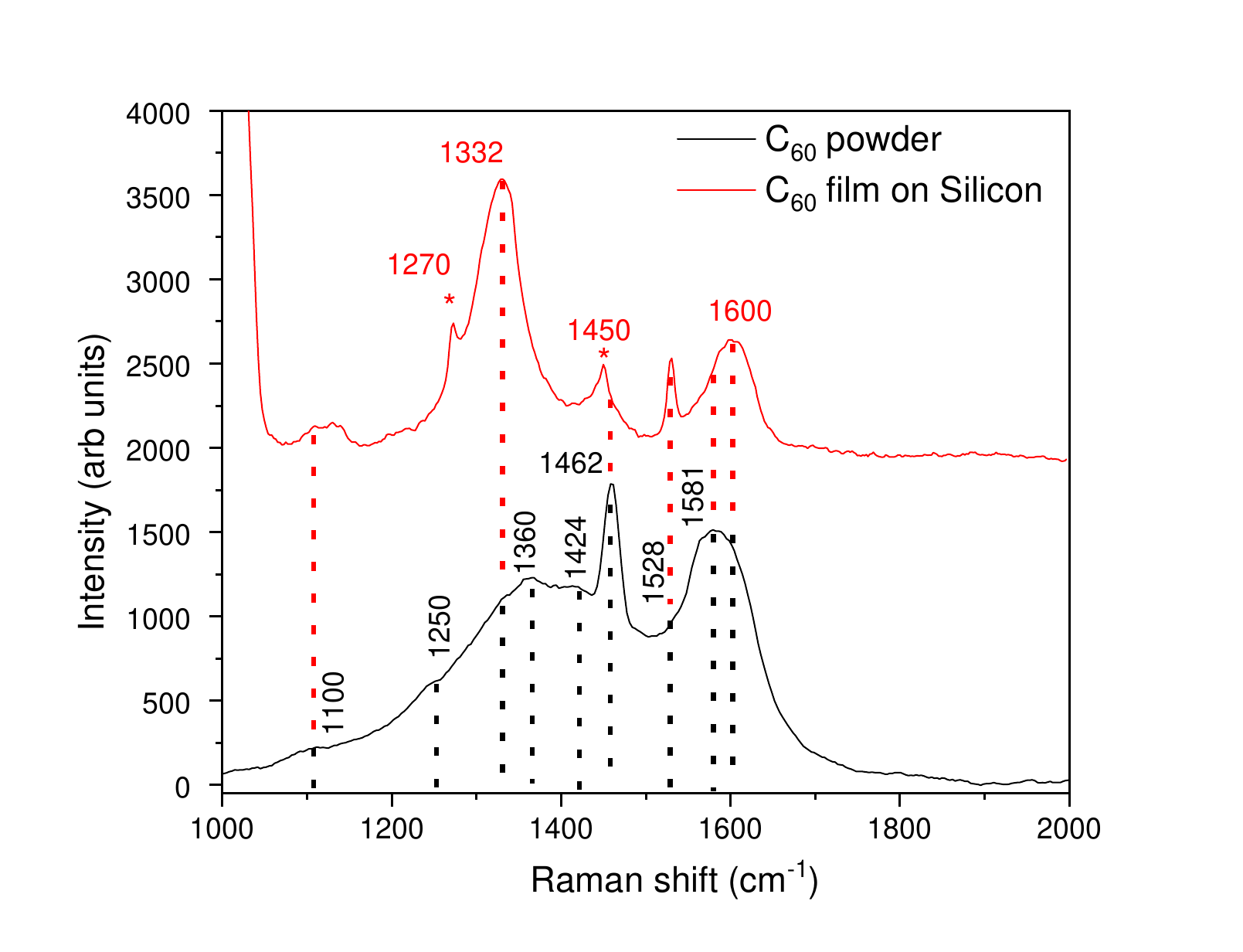}

\caption{Raman spectrum of C$_{60}$ powder (black line) and C$_{60}$ film deposited onto silicon substrate (red line). Peaks exclusive to the C$_{60}$ film spectrum are marked with asterisks, while the dotted lines (black and red) highlight peaks common to both C$_{60}$ powder and film.}
\label{RamanC60}

\end{figure}
\newpage
All treatments applied to the films to alter their properties and morphologies resulted in the development of numerous structures on the surface. One of the possible reasons behind these phenomena is the stress accumulated on the surface (and immediately under the surface) during the deposition, due to the immiscibility of the materials.  This stress is released almost instantaneously with the treatments, creating separation of the phases. The energy in excess, provided by the treatment is used for the growth of structures around the nucleation point. Depending on the available energy, more or less organized structures can form.

From the large area SEM, the sample that was subjected to annealing (Fig. \ref{fig_SEM}a) presents structures tending to the circular shape, with a strong nucleation of C$_{60}$ above the Co (or viceversa). Some structures show a main nucleation surrounded by a huge number of smaller nucleations (see Fig.S3 of SI for additional pictures). The sample exposed to $Ar^+$ beam irradiation (Fig. \ref{fig_SEM}b) displays a great number of very small separation spots, non-organized and spread randomly on the surface (see Fig. S4 of SI for additional pictures). The $C^+$ irradiation (Fig. \ref{fig_SEM}c) did not produce a particular separation on this films surface at least visible, but possible compositional changes in the C phase (see Fig. S5 of SI for additional pictures). The laser irradiation showed in Fig. \ref{fig_SEM}d produced large spots with C nucleation, but also microstructures with resemblance of long crystal structures (see Fig. S6 of SI for additional pictures).
The EDX mapping shows that the agglomerations of particles are formed by both constituents, with little more content of C in the center. Compared with the pristine sample there is the elemental distributions are more homogeneous, due to the release of existing stress. Additional EDX mapping are shown in SI Fig. S7-10.

A detailed Raman analysis of the pristine films of pure and hybrid materials was performed. Fig.\ref{RamanC60} shows the Raman spectra of the C$_{60}$ film deposited on the silicon substrate, compared to the vibrational fingerprint of pure C$_{60}$ powder used as a precursor in film preparation, obtained by excitement with a 532 nm laser wavelength. The Raman spectrum of the C$_{60}$ powder (black line) displays several Raman contributions between 1000 and 2000 $cm^{-1}$, closely matching values found in the literature \cite{dresselhaus1996raman, dorner2022effect}. Notably the Raman-active modes of $H_g$ symmetry are found at $\sim$1100 $cm^{-1}$ ($H_g$(5)), $\sim$1250 $cm^{-1}$ ($H_g$(6)), $\sim$1424 $cm^{-1}$($H_g$(7)), and 1581 $cm^{-1}$($H_g$(8)). In this Raman shift region, the peak at 1462 $cm^{-1}$ is assigned to the $A_g$(2) mode, known as “pentagonal pinch” since caused by the contraction of the pentagonal rings and the expansion of the hexagonal rings of carbon atoms. These peaks overlay the broader bands observed in the spectral ranges of 1330–1360 $cm^{-1}$ and 1560–1600 $cm^{-1}$, corresponding to the D and G bands, respectively, which are characteristic of all carbon-based materials under visible-range excitation \cite{ ferrari2001resonant}. Notably, laser heating during Raman measurements can facilitate the formation of additional carbon phases. 

\begin{figure}[h]
\begin{center}
\includegraphics[width=0.8\linewidth]{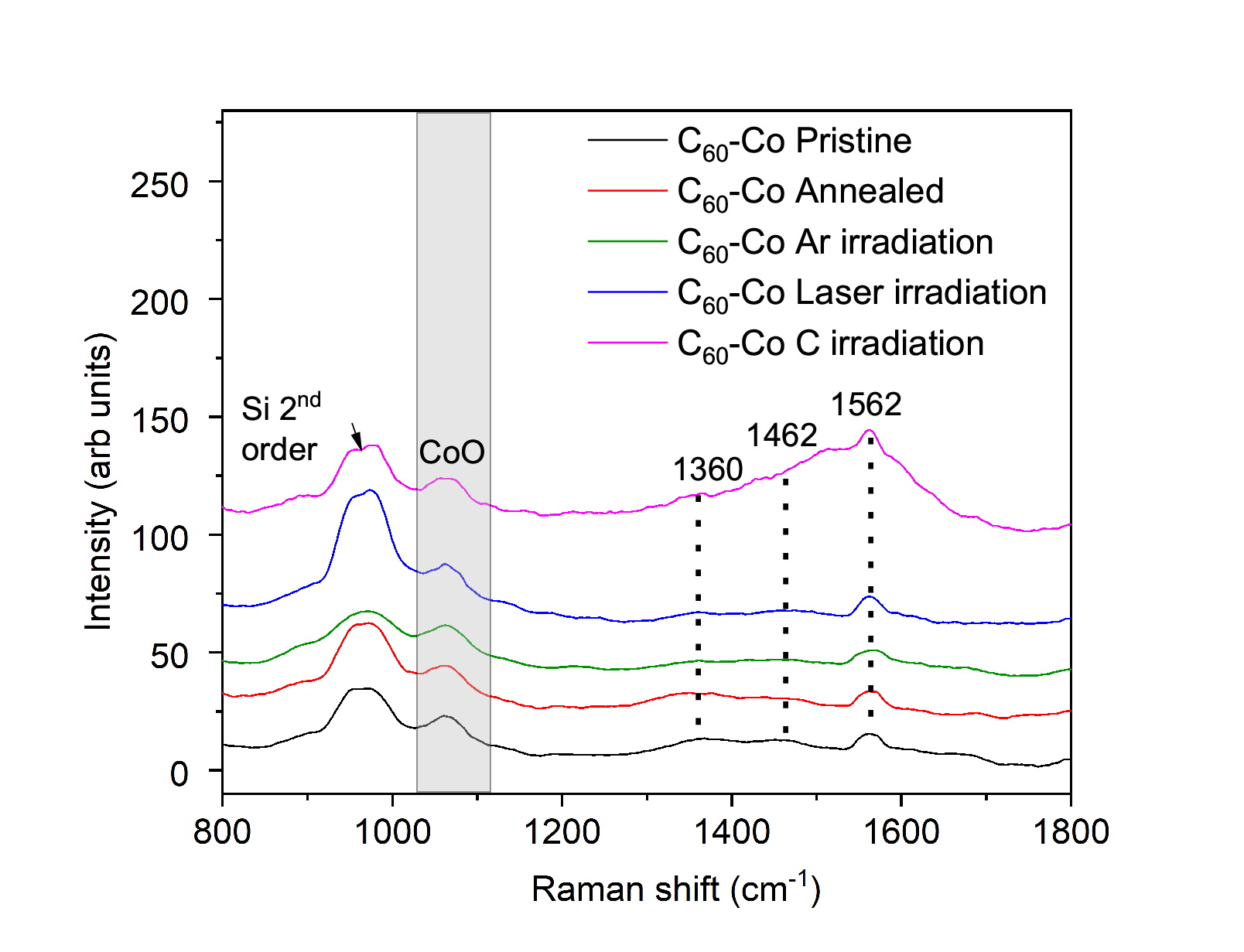} 
\end{center} 
\caption{Raman spectra of hybrid Co-C$_{60}$ films, as deposited (black line) onto silicon substrate and post-treatments: Co-C$_{60}$ film annealed (red line), Co-C$_{60}$ film Ar irradiated (green line), Co-C$_{60}$ film laser irradiated (blue line), and Co-C$_{60}$ film C irradiated (magenta line). The spectra are stacked in intensity for a better visualization.}
\label{Ramanfilms}

\end{figure}

The complete spectrum of C$_{60}$ powder, which includes the remaining four low-frequency $H_g$ symmetry modes and the purely radial $A_g$(1) mode at 496 $cm^{-1}$, is shown in Fig.S11 of the Supplementary Information. The Raman spectrum of C$_{60}$ film (red line in Fig.\ref{RamanC60}) shows a distinctive peak at 1450 $cm^{-1}$ (highlighted by asterisk), which attests a photo-transformed state of C$_{60}$ and the presence of polymer chains that, in the case of film, easily form during laser irradiation in Raman measurements \cite{dresselhaus1996raman}. Furthermore, the $A_g$(2) peak at 1462 $cm^{-1}$, characteristic of the C$_{60}$ powder, almost disappears in the fullerene film as it is overshadowed by the tail of the prominent peak at 1450 $cm^{-1}$. It is worthy to note the presence of the wide D band mainly peaked at 1332 $cm^{-1}$, typical of diamond-like carbon materials. The peak at 1270 $cm^{-1}$ (highlighted by asterisk) is ascribed to a combination of antisymmetric C-C interring stretching mode and antisymmetric bending mode of C-H groups that form on fullerene surface \cite{martin2015resonance}. The prominent and sharp peak tail that diminishes at 1065 $cm^{-1}$ corresponds to the second-order Raman peak of the Si substrate. Fig. \ref{Ramanfilms} displays the Raman spectra of hybrid Co-C$_{60}$ films, both pristine and post-treatment, within the 800–1800 $cm^{-1}$ spectral range. Notably, the characteristic $A_g$(2) peak at 1462 $cm^{-1}$ is significantly broadened and overlaps with the D and G bands (at 1360 and 1562 $cm^{-1}$, respectively), which are typical features of carbon materials. 
When the Co-C$_{60}$ is irradiated by carbon beam, a pronounced broad band appears beneath this peak, which is attributed to the formation of amorphous carbon. Furthermore, all spectra exhibit a broad band in the spectral range between 1030 and 1100 $cm^{-1}$ (grey rectangle in Fig. \ref{Ramanfilms}), attributed to the second-order two-phonon Raman scattering of the CoO phase \cite{li2016identification}. This spectral feature indicates cobalt oxidation induced in defective sites of cobalt during the co-deposition of hybrid films. Under laser heating during Raman measurements, the CoO phase readily transforms into the Co$_3$O$_4$ phase \cite{rivas2017thermodynamic}.
\begin{figure}[h]
\begin{minipage}[h]{0.48\linewidth}
\begin{center}
\includegraphics[width=1\linewidth]{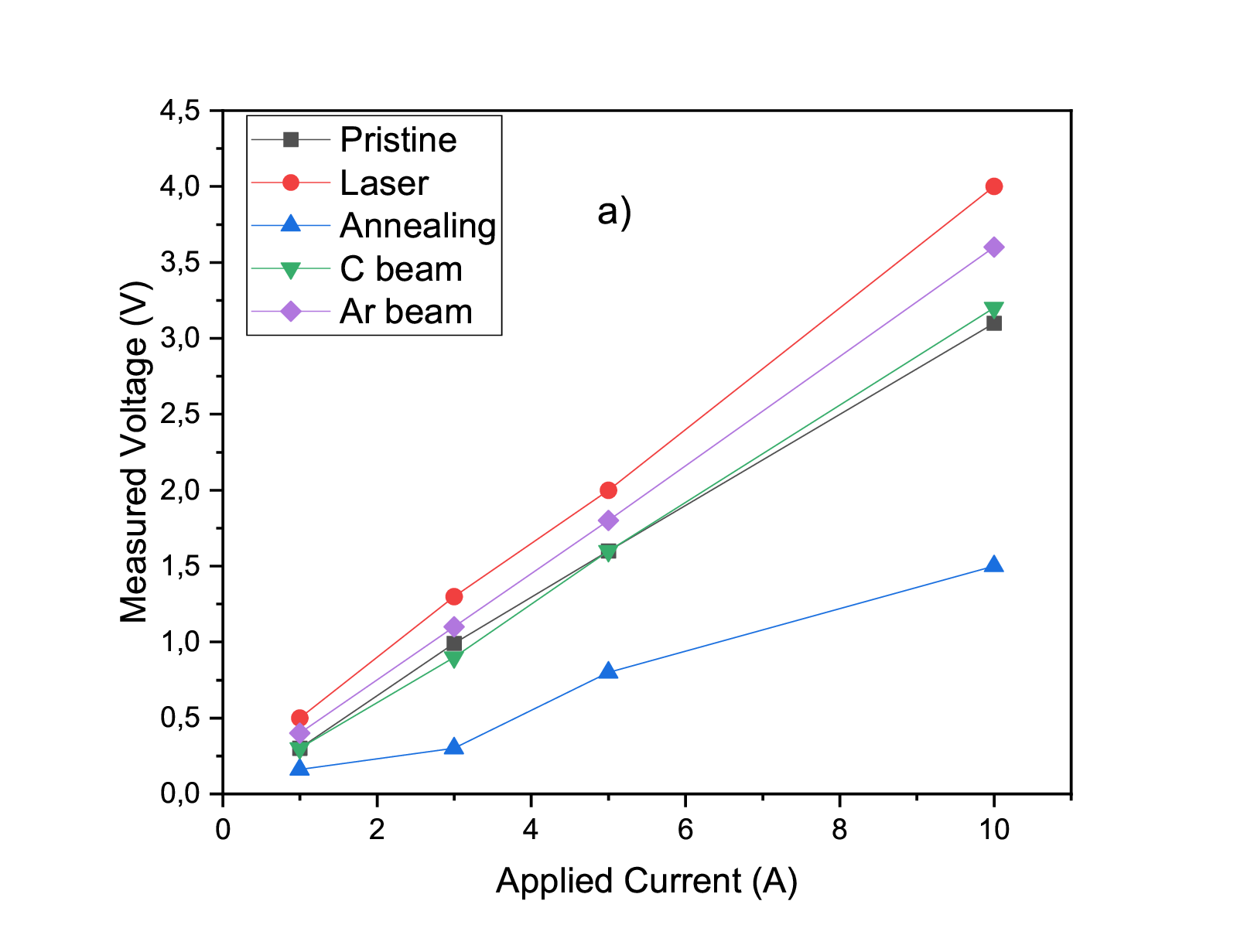} 
\label{fig6a}
\end{center} 
\end{minipage}
\begin{minipage}[h]{0.48\linewidth}
\begin{center}
\includegraphics[width=1\linewidth]{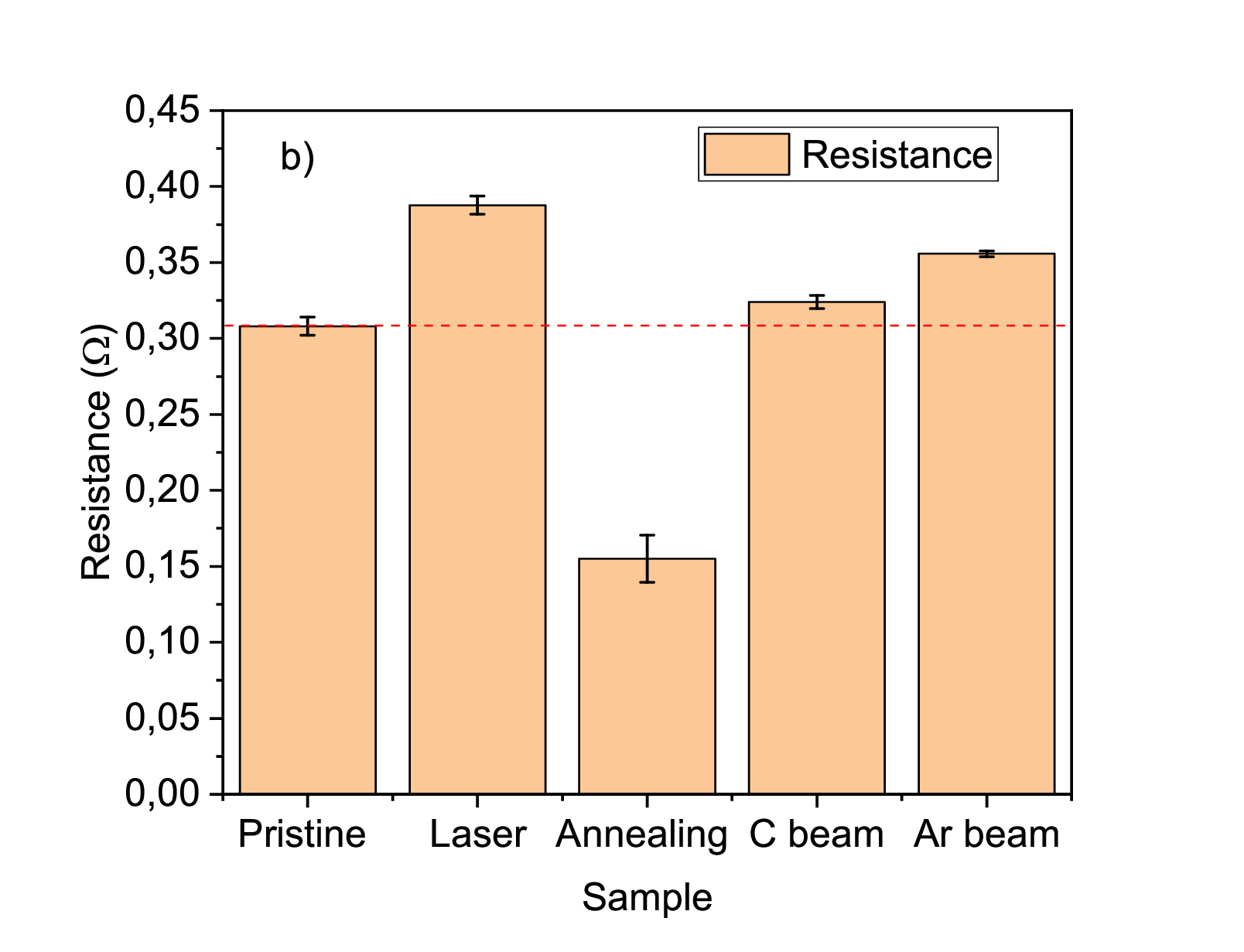} 
\label{fig6b}
\end{center} 
\end{minipage}
\caption{Results of galvanostatic measurement a) and calculated electrical resistance b), measured using 2-points method.}

\label{resistance}

\end{figure}

The results concerning the electrical resistance measurements performed on the films are shown in Fig. \ref{resistance}. Fig. \ref{resistance}a shows the galvanostatic measurements performed on the samples surface, using the setup described in "Material and Methods" section. The voltage measured with a nanovoltmeter between the electrodes that applied stabilized current to the films were recorded for several applied current values that ranged from 1 A to 10 A. Fig. \ref{resistance}b illustrates the resistance calculated by linear regression of the data in Fig. \ref{resistance}a. If we consider the pristine Co - C$_{60}$ film, it shows a resistance of 0.308 $\Omega$. It is a valid assumption considering this value as the reference for all the deposited films when the samples were all produced together during the same deposition. In this way, the measured variations in the resistance values are due only to the treatments to which they were exposed. Irradiation with a laser beam as well as bombardment with charged particles increased the resistance of the samples. In particular, laser irradiation leads to the highest resistance observed. An interesting result is instead the one referring to the sample subjected to annealing, for which the resistance value is 0.155 $\Omega$, almost half of the thin film in the pristine state. Probably, the annealing process produces large enucleations of C$_{60}$, separating the metal from the fullerene more clearly and the reduction in resistance observed after annealing is attributed to phase segregation and morphological reorganization, which likely promotes the formation of percolative Co-rich conduction pathways, even in the presence of partial cobalt oxidation.

\begin{figure}[h]
\begin{center}
\includegraphics[width=1\linewidth]{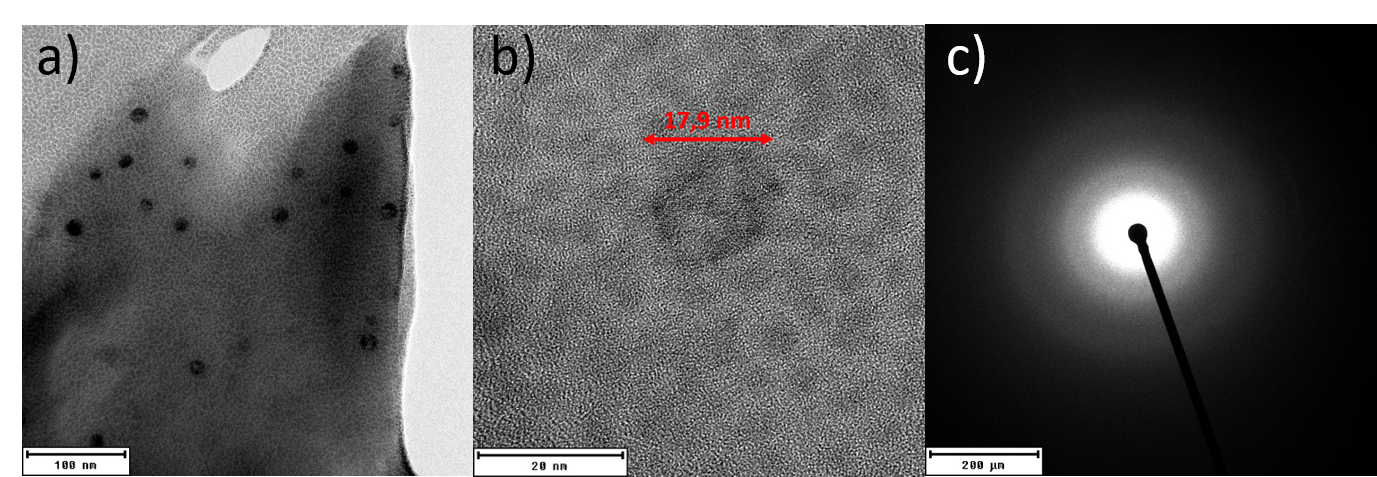} 
\end{center} 
\caption{TEM micrograph of annealed sample, magnification of fullerene NPs and SAED of area in picture.}
\label{TEM_annealed}

\end{figure}
\begin{figure}[h]
\begin{center}
\includegraphics[width=1\linewidth]{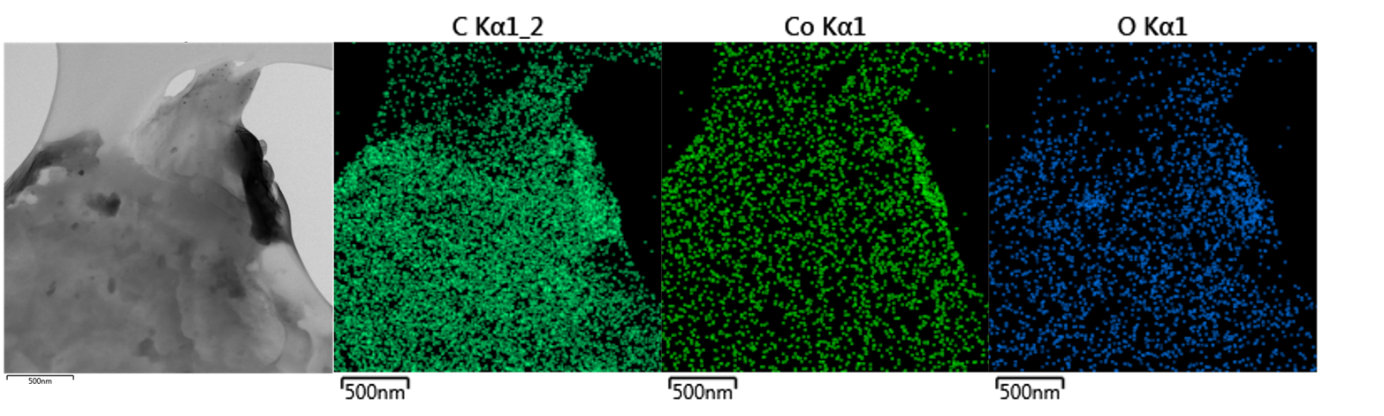} 
\end{center} 
\caption{Elemental mapping of area shown in Fig.\ref{TEM_annealed}}
\label{maptem}

\end{figure}

 
In order to investigate the origin of these particular electrical properties of the annealed samples, additional structural analyses were performed with TEM and the Selected Area Electron Diffraction (SAED) method on the grid sample. Fig.\ref{TEM_annealed}a) shows the morphology of selected area of modified film. It is possible to see agglomerations of self-assembled C$_{60}$ nanoparticles of the size of 17.9 nm (Fig.\ref{TEM_annealed}b). Instead, the SAED analysis allows us to measure the lattice parameters, crystal structure, and extent of crystallinity of nanoparticles from the diffraction technique, in which the sample is targeted with a parallel beam of high energy electrons.  In particular, Fig.\ref{TEM_annealed}c) illustrates that the film is in an amorphous state as expected.Finally, this area of the annealed grid sample was selected to perform elemental mapping also with the aim of understanding the elemental distribution of the elements (Fig.\ref{maptem}). As shown by the oxygen mapping, Cobalt effectively becomes oxidized, so the reduced electrical resistance has to be ascribable to a self-agglomeration of C$_{60}$ fullerene. Unfortunately, the carbon distribution is affected by the holey C film on which it is deposited.

\newpage
\section*{Discussion}

The systematic study performed in this work is focused on the preparation of Co–C$_{60}$ hybrid films under identical growth conditions and on their controlled modification through multiple external stimuli, enabling a direct comparison of distinct post-treatment pathways and their effects on nanostructure and functional properties. Well established ion beam analytical techniques, such as RBS and ERDA, have been used in the study to evaluate the film's original composition and thickness. ``As deposited'' and modified samples revealed several important morphological and structure evolutions based on the applied external stimuli. 

During the self-assembly of fullerene nanoparticles in a nonequilibrium solid-solution, C$_{60}$ molecules form a nanoparticle, finding the most advantageous combination of interactions between molecules with minimal free energy. Considering the efforts usually required to prepare the nanoparticle by pulverization, the possibility to obtain nanosized C$_{60}$ by dry physical method (in our case thermal evaporation) is truly an innovative result, considering that self-aggregation process of C$_{60}$ particles is purely physical, and is not associated with visible changing in phase crystal structure.
Interesting physical and structural properties were observed in the metal-C$_{60}$ phase without and with separation induced by external perturbations. SEM analyses performed on treated samples showed strong separations of the immiscible phases. The magnitude of such separations was different for each applied perturbation, as well as for the final shapes of formed structures. The nucleation produced by some treatments (e.g., the annealing process) produced a quasi-perfect circular germination of the material surrounded by smaller separation areas. These strong separations sometimes occur with neither the growth of carbon structures above the film nor with the formation of cobalt structures above the film. The study conducted using Raman spectroscopy allowed for the identification of the characteristic C$_{60}$ peak in the deposited film without Co, confirming the presence of the hollow structure of the fullerene. When combined with Co, there was a broadening of the characteristic $A_g$(2) peak, with an overlapping of D and G bands. In particular, it was observed that after carbon irradiation, a pronounced broad band appears beneath this peak due to the strong amorphization of carbon. The band attributed to the oxides of cobalt are also detected revealing a transformation of CoO in Co$_3$O$_4$ phase, as confirmed by Rivas-Murias and Salgueiriño in their previous work \cite{rivas2017thermodynamic}. 
Despite the presence of cobalt oxides, electrical resistance measurements revealed a non-monotonic response to the applied treatments. Ion and laser irradiation resulted in an increase in resistance, consistent with enhanced disorder, amorphization, and oxidation effects. In contrast, thermal annealing led to a pronounced reduction in the measured resistance in comparison to the pristine film. This behavior cannot be attributed to simple increase of Co amount, as oxidation of cobalt is clearly evidenced by Raman spectroscopy and elemental mapping.
Instead, the reduced resistance observed after annealing is to be attributed to morphology-driven reorganization and phase segregation within the hybrid film. TEM analysis of the annealed samples reveals the formation of self-assembled C$_{60}$ nanoparticles, indicating substantial molecular rearrangement and separation of the fullerene- and cobalt-rich regions. In such heterogeneous systems, cobalt is likely present in a mixed state comprising oxidized surface or interfacial regions surrounding Co-rich cores. In this case, electrical transport is governed by formation conduction pathways, where the macroscopic resistance is dominated by the connectivity of the most conductive regions rather than by the average phase composition.

Therefore, even in the presence of partial cobalt oxidation and intrinsically low fullerene conductivity, phase segregation induced by annealing can promote the formation of effective Co-rich conduction paths, resulting in a lower measured resistance. 
Overall, this study provides a comparative framework for understanding how different energy-delivery mechanisms affect Co–C$_{60}$ hybrid systems. The post-treatment–focused approach presented here offers a valuable strategy for tailoring the functional properties of metal–organic hybrid films and can be extended to other immiscible composite systems relevant to electronics, spintronics, and energy-related applications.
\ack{Sample text inserted for demonstration.}

\funding{The authors acknowledge the financial support from the MEYS CR, Project OP JAK FerrMion, No. CZ.02.01.01/00/22\_008/0004591.
The authors acknowledged also the support of Czech Academy of Science Mobility Plus Project, Grant No. JSPS-24-12 and JSPS Bilateral Program Number JPJSBP120242501. Measurements were carried out at the CANAM
infrastructure of the NPI CAS Rez under project LM 2015056. B.F. acknowledges the Italian Project PNRR “I-PHOQS - Integrated Infrastructure Initiative in Photonic and Quantum Sciences, CUP B53C22001750006.”. 
Part of this research was conducted within the activities of the RTD-A contract of S. Vasi co-funded by PON ‘Ricerca e Innovazione’ 2014-2020 (PON R\&I FSE-REACT EU), Azione IV.6 ‘Contratti di ricerca su tematiche Green’. C.C. acknowledges the financial support by the European Union – NextGeneration EU PNRR IR0000020 ECCSELLENT through NRRP – M4C2, Inv. 3.1 “Development of ECCSEL-R.I. Italian facilities: user access, services and long-term sustainability”}

\roles{G.C and K.T. conceived the experiment, G.C. and J.V. performed the samples preparation, G.C., E.S., K.T., Y.K., and R.M. conducted the experiment of sample modification, B.F., S.V.,C.C, A.M., R.M. and P.M. and J.N. performed the analysis, R.M., Y.K., S.V. analyzed the results, S.V., B.F. and G.C. manuscript preparation. All authors reviewed the manuscript.}

\data{Sample text inserted for demonstration.}

\suppdata{Sample text inserted for demonstration.}

\bibliographystyle{unsrt} 
\bibliography{sample} 

@Article{Scheele_2015,
  author    = {Scheele, M. and Brütting, W. and Schreiber, F.},
  journal   = {Physical Chemistry Chemical Physics},
  title     = {Coupled organic–inorganic nanostructures ({COIN)}},
  year      = {2015},
  issn      = {1463-9084},
  number    = {1},
  pages     = {97--111},
  volume    = {17},
  doi       = {10.1039/c4cp03094j},
  publisher = {Royal Society of Chemistry (RSC)},
}

@article{klingenberg1993oxygen,
  title={Oxygen adsorption and oxide formation on {C}o (1120)},
  author={Klingenberg, B and Grellner, F and Borgmann, D and Wedler, G},
  journal={Surface science},
  volume={296},
  number={3},
  pages={374--382},
  year={1993},
  publisher={Elsevier}
}

@article{dresselhaus1996raman,
  title={Raman scattering in fullerenes},
  author={Dresselhaus, MS and Dresselhaus, G and Eklund, PC},
  journal={Journal of Raman Spectroscopy},
  volume={27},
  number={3-4},
  pages={351--371},
  year={1996},
  publisher={Wiley Online Library}
}

@article{dorner2022effect,
  title={{Effect of fullerene C60 thermal and tribomechanical loading on Raman signals}},
  author={Dorner-Reisel, Annett and Ritter, Uwe and Moje, Jens and Freiberger, Emma and Scharff, Peter},
  journal={Diamond and Related Materials},
  volume={126},
  pages={109036},
  year={2022},
  publisher={Elsevier}
}

@article{martin2015resonance,
  title={{Resonance Raman spectroscopy and imaging of push--pull conjugated polymer--fullerene blends}},
  author={Martin, Eric JJ and B{\'e}rub{\'e}, Nicolas and Provencher, Fran{\c{c}}oise and C{\^o}t{\'e}, Michel and Silva, Carlos and Doorn, Stephen K and Grey, John K},
  journal={Journal of Materials Chemistry C},
  volume={3},
  number={23},
  pages={6058--6066},
  year={2015},
  publisher={Royal Society of Chemistry}
}

@article{vishnoi2020optical,
  title={{Optical and structural modifications of copper-fullerene nanocomposite thin films by 120 MeV Au ion irradiation}},
  author={Vishnoi, Ritu and Gupta, Satakshi and Dwivedi, Umesh Kumar and Singhal, Rahul},
  journal={Radiation Physics and Chemistry},
  volume={166},
  pages={108442},
  year={2020},
  publisher={Elsevier}
}

@article{sakai2007comparative,
  title={{Comparative study of structures and electrical properties in cobalt--fullerene mixtures by systematic change of cobalt content}},
  author={Sakai, Seiji and Naramoto, Hiroshi and Avramov, Pavel V and Yaita, Tsuyoshi and Lavrentiev, Vasily and Narumi, Kazumasa and Baba, Yuji and Maeda, Yoshihito},
  journal={Thin Solid Films},
  volume={515},
  number={20-21},
  pages={7758--7764},
  year={2007},
  publisher={Elsevier}
}

@article{rivas2017thermodynamic,
  title={{Thermodynamic CoO - Co3O4 crossover using Raman spectroscopy in magnetic octahedron-shaped nanocrystals}},
  author={Rivas-Murias, Beatriz and Salgueiri{\~n}o, Ver{\'o}nica},
  journal={Journal of Raman Spectroscopy},
  volume={48},
  number={6},
  pages={837--841},
  year={2017},
  publisher={Wiley Online Library}
}

@article{li2016identification,
  title={{Identification of cobalt oxides with Raman scattering and Fourier transform infrared spectroscopy}},
  author={Li, Yang and Qiu, Wenlan and Qin, Fan and Fang, Hui and Hadjiev, Viktor G and Litvinov, Dmitri and Bao, Jiming},
  journal={The Journal of Physical Chemistry C},
  volume={120},
  number={8},
  pages={4511--4516},
  year={2016},
  publisher={ACS Publications}
}

@article{ferrari2001resonant,
  title={{Resonant Raman spectroscopy of disordered, amorphous, and diamondlike carbon}},
  author={Ferrari, Andrea Carlo and Robertson, John},
  journal={Physical review B},
  volume={64},
  number={7},
  pages={075414},
  year={2001},
  publisher={APS}
}

@article{lin2021doping,
  title={{Doping graphene with substitutional mn}},
  author={Lin, Pin-Cheng and Villarreal, Renan and Achilli, Simona and Bana, Harsh and Nair, Maya N and Tejeda, Antonio and Verguts, Ken and De Gendt, Stefan and Auge, Manuel and Hofs{\"a}ss, Hans and others},
  journal={ACS nano},
  volume={15},
  number={3},
  pages={5449--5458},
  year={2021},
  publisher={ACS Publications}
}

@article{langer2022graphene,
  title={{Graphene lattices with embedded transition-metal atoms and tunable magnetic anisotropy energy: implications for spintronic devices}},
  author={Langer, Rostislav and Mustonen, Kimmo and Markevich, Alexander and Otyepka, Michal and Susi, Toma and B{\l}o{\'n}ski, Piotr},
  journal={ACS Applied Nano Materials},
  volume={5},
  number={1},
  pages={1562--1573},
  year={2022},
  publisher={ACS Publications}
}

@article{zhou2011adsorption,
  title={{Adsorption of gas molecules on transition metal embedded graphene: a search for high-performance graphene-based catalysts and gas sensors}},
  author={Zhou, Miao and Lu, Yun-Hao and Cai, Yong-Qing and Zhang, Chun and Feng, Yuan-Ping},
  journal={Nanotechnology},
  volume={22},
  number={38},
  pages={385502},
  year={2011},
  publisher={IOP Publishing}
}

@article{cortes2018fe,
  title={{Fe-doped graphene nanosheet as an adsorption platform of harmful gas molecules (CO, CO2, SO2 and H2S), and the co-adsorption in O2 environments}},
  author={Cort{\'e}s-Arriagada, Diego and Villegas-Escobar, Nery and Ortega, Daniela E},
  journal={Applied Surface Science},
  volume={427},
  pages={227--236},
  year={2018},
  publisher={Elsevier}
}

@article{zhang2020single,
  title={{Single-atom catalysts for electrocatalytic applications}},
  author={Zhang, Qiaoqiao and Guan, Jingqi},
  journal={Advanced Functional Materials},
  volume={30},
  number={31},
  pages={2000768},
  year={2020},
  publisher={Wiley Online Library}
}

@article{deng2016catalysis,
  title={{Catalysis with two-dimensional materials and their heterostructures}},
  author={Deng, Dehui and Novoselov, KS and Fu, Qiang and Zheng, Nanfeng and Tian, Zhongqun and Bao, Xinhe},
  journal={Nature nanotechnology},
  volume={11},
  number={3},
  pages={218--230},
  year={2016},
  publisher={Nature Publishing Group UK London}
}

@article{wang2018situ,
  title={{In situ formation of molecular Ni-Fe active sites on heteroatom-doped graphene as a heterogeneous electrocatalyst toward oxygen evolution}},
  author={Wang, Jiong and Gan, Liyong and Zhang, Wenyu and Peng, Yuecheng and Yu, Hong and Yan, Qingyu and Xia, Xinghua and Wang, Xin},
  journal={Science advances},
  volume={4},
  number={3},
  pages={eaap7970},
  year={2018},
  publisher={American Association for the Advancement of Science}
}

@Article{Goiri_2015,
  author    = {Goiri, Elizabeth and Borghetti, Patrizia and El‐Sayed, Afaf and Ortega, J. Enrique and de Oteyza, Dimas G.},
  journal   = {Advanced Materials},
  title     = {{Multi‐Component Organic Layers on Metal Substrates}},
  year      = {2015},
  issn      = {1521-4095},
  month     = dec,
  number    = {7},
  pages     = {1340--1368},
  volume    = {28},
  doi       = {10.1002/adma.201503570},
  publisher = {Wiley},
}

@Article{Cui_2016,
  author    = {Cui, Yuanjing and Li, Bin and He, Huajun and Zhou, Wei and Chen, Banglin and Qian, Guodong},
  journal   = {Accounts of Chemical Research},
  title     = {{Metal–Organic Frameworks as Platforms for Functional Materials}},
  year      = {2016},
  issn      = {1520-4898},
  month     = feb,
  number    = {3},
  pages     = {483--493},
  volume    = {49},
  doi       = {10.1021/acs.accounts.5b00530},
  publisher = {American Chemical Society (ACS)},
}

@Article{Sosa_2018,
  author    = {Sosa, Joshua D. and Bennett, Timothy F. and Nelms, Katherine J. and Liu, Brandon M. and Tovar, Roberto C. and Liu, Yangyang},
  journal   = {Crystals},
  title     = {{Metal–Organic Framework Hybrid Materials and Their Applications}},
  year      = {2018},
  issn      = {2073-4352},
  month     = aug,
  number    = {8},
  pages     = {325},
  volume    = {8},
  doi       = {10.3390/cryst8080325},
  publisher = {MDPI AG},
}

@Article{Devkota_2016,
  author    = {Devkota, Jagannath and Geng, Rugang and Subedi, Ram Chandra and Nguyen, Tho Duc},
  journal   = {{Advanced Functional Materials}},
  title     = {Organic Spin Valves: A Review},
  year      = {2016},
  issn      = {1616-3028},
  month     = mar,
  number    = {22},
  pages     = {3881--3898},
  volume    = {26},
  doi       = {10.1002/adfm.201504209},
  publisher = {Wiley},
}

@Article{Socol_2020,
  author    = {Socol, Marcela and Preda, Nicoleta and Costas, Andreea and Borca, Bogdana and Popescu-Pelin, Gianina and Mihailescu, Andreea and Socol, Gabriel and Stanculescu, Anca},
  journal   = {Nanomaterials},
  title     = {{Thin Films Based on {C}obalt {P}hthalocyanine:{C}60 Fullerene:{Z}n{O} Hybrid Nanocomposite Obtained by Laser Evaporation}},
  year      = {2020},
  issn      = {2079-4991},
  month     = mar,
  number    = {3},
  pages     = {468},
  volume    = {10},
  doi       = {10.3390/nano10030468},
  publisher = {MDPI AG},
}

@Article{Song_2019,
  author    = {Song, Xiaoyu and Wang, Xinyue and Li, Yusen and Zheng, Chengzhi and Zhang, Bowen and Di, Chong‐an and Li, Feng and Jin, Chao and Mi, Wenbo and Chen, Long and Hu, Wenping},
  journal   = {Angewandte Chemie International Edition},
  title     = {{2D Semiconducting Metal–Organic Framework Thin Films for Organic Spin Valves}},
  year      = {2019},
  issn      = {1521-3773},
  month     = nov,
  number    = {3},
  pages     = {1118--1123},
  volume    = {59},
  doi       = {10.1002/anie.201911543},
  publisher = {Wiley},
}

@Article{Shao_2023,
  author    = {Shao, Zhichao and Chen, Junshuai and Gao, Kexin and Xie, Qiong and Xue, Xiaojing and Li, Xue and Hou, Hongwei and Mi, Liwei},
  journal   = {Chemical Engineering Journal},
  title     = {{A high-spintronic helix metal-organic chain as a high-output triboelectric nanogenerator material for self-powered anticorrosion}},
  year      = {2023},
  issn      = {1385-8947},
  month     = jan,
  pages     = {140865},
  volume    = {455},
  doi       = {10.1016/j.cej.2022.140865},
  publisher = {Elsevier BV},
}

@Article{Otero_2017,
  author    = {Otero, R. and Vázquez de Parga, A.L. and Gallego, J.M.},
  journal   = {Surface Science Reports},
  title     = {{Electronic, structural and chemical effects of charge-transfer at organic/inorganic interfaces}},
  year      = {2017},
  issn      = {0167-5729},
  month     = jul,
  number    = {3},
  pages     = {105--145},
  volume    = {72},
  doi       = {10.1016/j.surfrep.2017.03.001},
  publisher = {Elsevier BV},
}

@Article{Yoshikawa_2009,
  author    = {Yoshikawa, Genki and Tsuruma, Yuki and Ikeda, Susumu and Saiki, Koichiro},
  journal   = {Advanced Materials},
  title     = {{Noble Metal Intercalated Fullerene Fabricated by Low‐Temperature Co‐deposition}},
  year      = {2009},
  issn      = {1521-4095},
  month     = dec,
  number    = {1},
  pages     = {43--46},
  volume    = {22},
  doi       = {10.1002/adma.200900921},
  publisher = {Wiley},
}

@Article{Konarev_2003,
  author    = {Konarev, Dmitri V. and Khasanov, Salavat S. and Saito, Gunzi and Lyubovskaya, Rimma N. and Yoshida, Yukihiro and Otsuka, Akihiro},
  journal   = {Chemistry - A European Journal},
  title     = {{The Interaction of C60, C70, and C60(CN)2 Radical Anions with Cobalt(II) Tetraphenylporphyrin in Solid Multicomponent Complexes}},
  year      = {2003},
  issn      = {1521-3765},
  month     = aug,
  number    = {16},
  pages     = {3837--3848},
  volume    = {9},
  doi       = {10.1002/chem.200204470},
  publisher = {Wiley},
}

@Article{Tanaka_2002,
  author    = {Tanaka, H and Marumoto, K and Kuroda, S and Ishii, T and Kanehama, R and Aizawa, N and Matsuzaka, H and Sugiura, K and Miyasaka, H and Kodama, T and Kikuchi, K and Ikemoto, I and Yamashita, M},
  journal   = {Journal of Physics: Condensed Matter},
  title     = {{Electron spin resonance studies of Co(tbp)·C60single crystal}},
  year      = {2002},
  issn      = {1361-648X},
  month     = apr,
  number    = {15},
  pages     = {3993--4000},
  volume    = {14},
  doi       = {10.1088/0953-8984/14/15/313},
  publisher = {IOP Publishing},
}

@Article{Vacik_2010,
  author    = {Vacik, J. and Lavrentiev, V. and Novotna, K. and Bacakova, L. and Lisa, V. and Vorlicek, V. and Fajgar, R.},
  journal   = {Diamond and Related Materials},
  title     = {{Fullerene (C60)–transitional metal (Ti) composites: Structural and biological properties of the thin films}},
  year      = {2010},
  issn      = {0925-9635},
  month     = feb,
  number    = {2–3},
  pages     = {242--246},
  volume    = {19},
  doi       = {10.1016/j.diamond.2009.10.016},
  publisher = {Elsevier BV},
}

@Article{Bolokang_2012,
  author    = {Bolokang, A.S. and Phasha, M.J. and Camagu, S.T. and Motaung, D.E. and Bhero, S.},
  journal   = {International Journal of Refractory Metals and Hard Materials},
  title     = {{Effect of thermal treatment on mechanically milled cobalt powder}},
  year      = {2012},
  issn      = {0263-4368},
  month     = mar,
  pages     = {258--262},
  volume    = {31},
  doi       = {10.1016/j.ijrmhm.2011.11.010},
  publisher = {Elsevier BV},
}

@Article{Gupta_2023,
  author    = {Gupta, Suraj and Fernandes, Rohan and Patel, Rupali and Spreitzer, Matjaž and Patel, Nainesh},
  journal   = {Applied Catalysis A: General},
  title     = {{A review of cobalt-based catalysts for sustainable energy and environmental applications}},
  year      = {2023},
  issn      = {0926-860X},
  month     = jul,
  pages     = {119254},
  volume    = {661},
  doi       = {10.1016/j.apcata.2023.119254},
  publisher = {Elsevier BV},
}

@Article{Patel_2023,
  author    = {Patel, Gauravkumar and Ganss, Fabian and Salikhov, Ruslan and Stienen, Sven and Fallarino, Lorenzo and Ehrler, Rico and Gallardo, Rodolfo A. and Hellwig, Olav and Lenz, Kilian and Lindner, Jürgen},
  journal   = {Physical Review B},
  title     = {{Structural and magnetic properties of thin cobalt films with mixed hcp and fcc phases}},
  year      = {2023},
  issn      = {2469-9969},
  month     = nov,
  number    = {18},
  pages     = {184429},
  volume    = {108},
  doi       = {10.1103/physrevb.108.184429},
  publisher = {American Physical Society (APS)},
}

@article{joshi2021fabrication,
  title={{Fabrication and applications of fullerene-based metal nanocomposites: A review}},
  author={Joshi, Gaurang and Mehta, Kush P and others},
  journal={Journal of Materials Research},
  volume={36},
  number={1},
  pages={114--128},
  year={2021},
  publisher={Springer}
}

@Article{Kaushik_2022,
  author    = {Kaushik, Sonia and Khanderao, Avinash G. and Gupta, Pooja and Raghavendra Reddy, V. and Kumar, Dileep},
  journal   = {Materials Science and Engineering: B},
  title     = {{Growth of ultra-thin Cobalt on fullerene (C60) thin-film: in-situ investigation under UHV conditions}},
  year      = {2022},
  issn      = {0921-5107},
  month     = oct,
  pages     = {115911},
  volume    = {284},
  doi       = {10.1016/j.mseb.2022.115911},
  publisher = {Elsevier BV},
}

@article{karczmarska2022carbon,
  title={{Carbon-supported noble-metal nanoparticles for catalytic applications—a review}},
  author={Karczmarska, Agnieszka and Adamek, Micha{\l} and El Houbbadi, Sara and Kowalczyk, Pawe{\l} and Laskowska, Magdalena},
  journal={Crystals},
  volume={12},
  number={5},
  pages={584},
  year={2022},
  publisher={MDPI}
}

@article{varshney1999superconductivity,
  title={{Superconductivity in alkali metal doped fullerenes (K3C60): a phonon mechanism}},
  author={Varshney, Dinesh and Varshney, Meenu and Singh, RK and Mishra, Raghvendra},
  journal={Journal of Physics and Chemistry of Solids},
  volume={60},
  number={5},
  pages={579--585},
  year={1999},
  publisher={Elsevier}
}

@article{takeya2013preparation,
  title={{Preparation and superconductivity of potassium-doped fullerene nanowhiskers}},
  author={Takeya, Hiroyuki and Kato, Ryoei and Wakahara, Takatsugu and Miyazawa, Kun’ichi and Yamaguchi, Takahide and Ozaki, Toshinori and Okazaki, Hiroyuki and Takano, Yoshihiko},
  journal={Materials Research Bulletin},
  volume={48},
  number={2},
  pages={343--345},
  year={2013},
  publisher={Elsevier}
}

@article{vacik2016laser,
  title={{Laser-induced periodic surface structure in nickel-fullerene composites}},
  author={Vacik, J and Lavrentiev, V and Havranek, V and Horak, P and Hnatowicz, V and Fajgar, R},
  journal={Radiation Effects and Defects in Solids},
  volume={171},
  number={1-2},
  pages={154--160},
  year={2016},
  publisher={Taylor \& Francis}
}

@inproceedings{vacik2009hybridization,
  title={{Hybridization And Modification Of The Ni/C60 Composites}},
  author={Vacik, J and Lavrentiev, V and Hnatowicz, V and Vorlicek, V and Naramoto, H},
  booktitle={AIP Conference Proceedings},
  volume={1099},
  number={1},
  pages={553--556},
  year={2009},
  organization={American Institute of Physics}
}

@article{lavrentiev2020structure,
  title={{Structure assembly regularities in vapour-deposited gold--fullerene mixture films}},
  author={Lavrentiev, V and Motylenko, M and Barchuk, M and Schimpf, C and Lavrentieva, I and Pokorn{\`y}, J and R{\"o}der, C and Vacik, J and Dejneka, A and Rafaja, D},
  journal={Nanoscale Advances},
  volume={2},
  number={4},
  pages={1542--1550},
  year={2020},
  publisher={Royal Society of Chemistry}
}

@article{lavrentiev2021tuneable,
  title={{Tuneable interplay of plasmonic and molecular excitations in self-assembled silver-fullerene nanocomposites}},
  author={Lavrentiev, Vasily and Chvostova, Dagmar and Pokorny, Jan and Lavrentieva, Inna and Vacik, Jiri and Dejneka, Alexandr},
  journal={Carbon},
  volume={184},
  pages={34--42},
  year={2021},
  publisher={Elsevier}
}

@article{lavrentiev2024room,
  title={{Room temperature excitonic coupling in self-assembled copper--Fullerene hybrid films exposed to ambient air}},
  author={Lavrentiev, Vasily and Chvostova, Dagmar and Klementova, Mariana and Kuldova, Karla and de Prado, Esther and Vacik, Jiri and Lavrentieva, Inna and Dejneka, Alexandr},
  journal={Carbon},
  volume={226},
  pages={119230},
  year={2024},
  publisher={Elsevier}
}

@article{lee2002excellent,
  title={{Excellent magnetic properties of fullerene encapsulated ferromagnetic nanoclusters}},
  author={Lee, GH and Huh, SH and Jeong, JW and Ri, H-C},
  journal={Journal of magnetism and magnetic materials},
  volume={246},
  number={3},
  pages={404--411},
  year={2002},
  publisher={Elsevier}
}

@article{zalibera2021metallofullerene,
  title={{Metallofullerene photoswitches driven by photoinduced fullerene-to-metal electron transfer}},
  author={Zalibera, Michal and Ziegs, Frank and Schiemenz, Sandra and Dubrovin, Vasilii and Lubitz, Wolfgang and Savitsky, Anton and Deng, Shihu HM and Wang, Xue-Bin and Avdoshenko, Stanislav M and Popov, Alexey A},
  journal={Chemical Science},
  volume={12},
  number={22},
  pages={7818--7838},
  year={2021},
  publisher={Royal Society of Chemistry}
}

@article{kuznetsov2012magnetic,
  title={{Magnetic properties of endohedral complexes Co5@ Cn depending upon the size and symmetry of fullerenes as well as orientation of cobalt cluster}},
  author={Kuznetsov, Andrew},
  journal={Computational Materials Science},
  volume={54},
  pages={204--207},
  year={2012},
  publisher={Elsevier}
}

@article{ceccio2024study,
  title={{Study of thin film composites based on LiCoO2 and C60 using neutron depth profiling and atomic force microscopy}},
  author={Ceccio, Giovanni and Vacik, Jiri and Lavrentiev, Vasyl and Tomandl, Ivo and Miksova, Romana and Takahashi, Kazumasa},
  journal={Journal of Radioanalytical and Nuclear Chemistry},
  volume={333},
  number={12},
  pages={6687--6697},
  year={2024},
  publisher={Springer}
}

@article{vacik2024study,
  title={{Study of surface morphology of Ag thin films prepared by sputtering and irradiation with keV Ar ion beam}},
  author={Vacik, J and Ceccio, G and Lavrentiev, V and Miksova, R and Havranek, V and Pleskunov, P and Cannav{\`o}, A},
  journal={Radiation Effects and Defects in Solids},
  volume={179},
  number={1-2},
  pages={136--145},
  year={2024},
  publisher={Taylor \& Francis}
}

@inproceedings{mayer1999simnra,
  title={{SIMNRA, a simulation program for the analysis of NRA, RBS and ERDA}},
  author={Mayer, Matej},
  booktitle={AIP conference proceedings},
  volume={475},
  number={1},
  pages={541--544},
  year={1999},
  organization={American Institute of Physics}
}

\end{document}